\documentclass[]{aastex}
\usepackage{emulateapj5}
\usepackage{onecolfloat}
\usepackage{graphicx} 
\usepackage{fancyheadings} 
\usepackage{ulem}
\usepackage{rotating}
\usepackage{lscape}

\newcommand{\qon}{BR0951$-$04}
\newcommand{\qoo}{PSS0132+13}
\newcommand{\qof}{PSS1443+27}
\newcommand{\zphot}{z_{phot}}
\newcommand{\zabs}{z_{abs}}

\newcommand{\Lya}{Ly$\alpha$ }
\newcommand{\lya}{Ly$\alpha$ }

\newcommand{\cm}[1]{\, {\rm cm^{#1}}}
\newcommand{\N}[1]{{N({\rm #1})}}

\newcommand{\rAA}{{\AA \enskip}}
\newcommand{\sci}[1]{{\rm \; \times \; 10^{#1}}}

\begin{document}

\twocolumn[%
\submitted{Accepted to the Astronomical Journal: Jan 22, 2002}

\title{GALAXIES ASSOCIATED WITH $z \sim 4$ DAMPED \lya SYSTEMS
I. IMAGING AND PHOTOMETRIC SELECTION }

\author{JASON X. PROCHASKA\altaffilmark{1,2,3},
ERIC GAWISER\altaffilmark{4},
ARTHUR M. WOLFE\altaffilmark{1,4},
ANDREAS QUIRRENBACH\altaffilmark{1,4},
KENNETH M. LANZETTA\altaffilmark{5},
HSIAO-WEN CHEN\altaffilmark{2},
JEFF COOKE\altaffilmark{4},
\& NORIAKI YAHATA \altaffilmark{5}}

\begin{abstract} 

This paper describes the acquisition and analysis of imaging data
for the identification of galaxies associated with $z \sim 4$ damped
\lya systems.  We present deep $BRI$ images of three fields known to
contain four $z \sim 4$ damped systems.  We discuss the reduction
and calibration of the data, detail the color criteria used
to identify $z \sim 4$ galaxies, and present a photometric redshift
analysis to complement the color selection.  We have found no galaxy
candidates closer to the QSO than $7''$ which could be responsible for
the damped \lya systems.  Assuming that at least one of the galaxies is
not directly beneath the QSO, we set an upper limit on this damped \lya
system of $L < L_{LBG}^*/4$.  Finally, we have established a web site to
release these imaging data to the 
public (http://kingpin.ucsd.edu/$\sim$dlaimg).

\end{abstract}
]

\altaffiltext{1}{Visiting Astronomer, W.M. Keck Telescope.
The Keck Observatory is a joint facility of the University
of California and the California Institute of Technology.}
\altaffiltext{2}{The Observatories of the Carnegie Institute of Washington,
813 Santa Barbara St., Pasadena, CA 91101}
\altaffiltext{3}{Hubble Fellow}
\altaffiltext{4}{Department of Physics, and Center for Astrophysics and 
Space Sciences, University of California, San Diego, C--0424, La Jolla, 
CA 92093--0424}
\altaffiltext{5}{Department of Physics and Astronomy, State University
of New York at Stony Brook, Stony Brook, NY, 11794--3800}

\pagestyle{fancyplain}
\lhead[\fancyplain{}{\thepage}]{\fancyplain{}{PROCHASKA ET AL.}}
\rhead[\fancyplain{}{Galaxies Associated with $z \sim 4$ Damped \Lya Systems I.}]{\fancyplain{}{\thepage}}
\setlength{\headrulewidth=0pt}
\cfoot{}

\section{INTRODUCTION}

In this era of large telescopes, it is almost routine to identify
galaxies at high redshift.  Deep narrow band imaging
\citep[e.g.][]{hu98} and the Lyman break technique \citep{steidel93},
among other approaches, have led to the discovery of over 1000
$z > 2$ galaxies where just 10 years ago only a handful were known.
Unfortunately, because these techniques are magnitude limited, 
they are primarily sensitive to the high luminosity tail of the
protogalactic distribution.  For example, the
Lyman break galaxies exhibit a two-point correlation function which suggests
they are massive galaxies associated with significant large-scale 
structures \citep{adlb98} that will evolve into the 
most massive galaxies today.  To study the 
formation of typical galaxies, therefore, one must 
circumvent this luminosity selection. 

Even before the recent outburst of high redshift
galaxy identifications, researchers
had already discovered a significant
population of high $z$ protogalaxies in absorption: 
the damped \lya systems \citep[e.g.][]{wol86}.
Because the selection of these 
galaxies is based on H~I gas cross-section, they are predicted
to sample the bulk of the galactic mass distribution \citep{kau96} and
therefore provide a broader picture of galaxy formation in the early universe.
Attempts to identify in emission
the galaxies which give rise to damped systems, however, have met with
limited success.  Djorgovski and collaborators \citep{djg96,djg97}
have made a few discoveries, but their success rate is uncertain.
More recently, \cite{warren01} have analysed deep HST images of
fields around quasars known to exhibit damped \lya systems and have
identified small, relatively faint galaxies close to the quasar
claimed to be the damped \lya systems.
In both of these cases, the surveys have
focused on galactic candidates within a 
few arcseconds of the background quasar.  
Although this is an important aspect of damped \lya research,
there are unique advantages to extending the galaxy surveys to 
a much larger area. 
In particular, one can estimate the mass of the damped systems -- 
and therefore the bulk of the high $z$ protogalactic population -- by
measuring their clustering properties \citep{gawiser01a}. 

This paper describes the observational and analysis techniques we have
implemented to select high redshift galaxy candidates in deep $BRI$
images of three fields containing four known damped \lya systems.
It marks the first step in an
observing program designed to search in emission
for the galaxies giving rise to or clustered with damped \lya 
systems at high redshift.
The present observations focus on fields with known
$\zabs \sim 4$ damped \lya systems, i.e.\ the galaxies associated
with these systems will be $B$-band dropouts.  
While follow-up observations of $z \sim 4$ galaxies are considerably
more difficult than at $z \sim 3$ \citep{steidel99}, we have been
limited by the absence of a $U$-band imager at the Keck Observatories.
Table~\ref{tab:dla} lists the known properties
of the four damped \lya systems from the three quasar fields
described in this paper.  Each 
system exhibits a neutral hydrogen column density
which exceeds the statistical survey threshold
$(\N{HI} = 2 \sci{20} \cm{-2}$) established by \cite{wol86}.  
For two of the quasars we have obtained high resolution spectroscopy
\citep{pro99,pro00,pro01} with HIRES \citep{vogt94} on the Keck~I telescope
and have measured their [Fe/H] metallicity.
The damped system toward {\qof} is remarkable for exhibiting by far the 
highest metallicity of any $z>3$ damped \lya system to date while the 
$z = 4.20$ system toward {\qon} has among the lowest metallicity.

\begin{table}[ht]
\begin{center}
\caption{{\sc DAMPED LYA SYSTEMS
\label{tab:dla}}}
\begin{tabular}{lcccc}
\tableline
\tableline
Quasar & $z_{QSO}$ & $z_{DLA}$ & 
$\N{HI}$\tablenotemark{a} & [Fe/H] \\
\tableline
PSS0132+1341  & 4.15 & 3.93 & 20.3 & \\
BR0951$-$0450 & 4.37 & 4.20 & 20.4 & $< -2.6$\tablenotemark{b} \\
              &      & 3.86 & 20.6 & $-2.1$\tablenotemark{b} \\
PSS1443+2724  & 4.41 & 4.22 & 20.8 & $-1.0$\tablenotemark{c} \\
\tableline
\end{tabular}
\tablenotetext{a}{\cite{storr00}}
\tablenotetext{b}{\cite{pro99}}
\tablenotetext{c}{\cite{pro00,pro01}}
\end{center}
\end{table}

First results on one of these fields (\qon) have been presented elsewhere
\citep{gawiser01a} and a companion paper \citep[][Paper II]{gawiser01b} 
describes
the reduction and analysis of the multi-slit spectroscopy of all three fields.
In $\S$~\ref{sec-obsred} we discuss the imaging observations and reduction
process, the photometric calibration is described in $\S$~\ref{sec-photcalib},
we present the photometry and astrometry for all detected objects in 
$\S$~\ref{sec:photom}, $\S$~\ref{sec:results} presents the results, and
$\S$~\ref{sec:summary} gives a brief summary.
Unless otherwise noted, all magnitudes have been placed on the $AB_{95}$ scale 
by offsetting Johnson-Cousins magnitudes: $B_{AB} = B_{JC} - 0.11;
R_{AB} = R_{JC} + 0.20; I_{AB} = I_{JC} + 0.45$ \citep{fuk96}.

\begin{table*}
\begin{center}
\caption{{\sc JOURNAL OF OBSERVATIONS \label{tab:obs}}}
\begin{tabular}{lcccccc}
\tableline
\tableline
QSO Field & UT & Filter
& Eff Exp & FWHM & SB$^{1\sigma}$ &
SB$^{3\sigma}$ \\
& & & (s) & ($''$) & (Mag$_{AB}$/$\square ''$) & (Mag$_{AB}$/FWHM$^2$) \\
\tableline
PSS0132+1341  & Jan 13 1999 & B & 3750 & 1.05 & 29.5 & 28.3 \\
              &             & R & 1680 & 0.85 & 28.8 & 27.8 \\
              &             & I & 1080 & 0.58 & 28.0 & 27.4 \\
BR0951$-$0450 & Jan 13 1999 & B & 4500 & 0.85 & 29.8 & 28.8 \\
              &             & R & 1620 & 0.75 & 29.0 & 28.1 \\
              &             & I & 1560 & 0.63 & 28.1 & 27.4 \\
PSS1443+2724  & Mar 23 1999 & B & 3600 & 1.08 & 29.7 & 28.4 \\
              &             & R & 1200 & 0.80 & 28.8 & 27.9 \\
              &             & I & 750  & 1.02 & 27.8 & 26.6 \\
\tableline
\end{tabular}
\end{center}
\end{table*}

\section{OBSERVATIONS AND DATA REDUCTION}
\label{sec-obsred}

For this initial study, 
the specific goal of the imaging program was to acquire deep enough
$BRI$ images to identify all $B$-band dropouts ($B-R > 2$) 
bright enough to obtain
follow-up spectroscopy with LRIS \citep{oke94} at the
Keck observatory (i.e.\ $I < 25$).  
Table~\ref{tab:obs} provides a journal of our imaging observations
taken over the course of two nights in early 1999
with LRIS on the Keck~II telescope.
Conditions on both nights were photometric. 
The Tektronix 2048x2048 CCD (0.213$''$/pixel)
was read out in dual-amp mode with a read-out noise 
of 7 e$^-$ and a gain of $\approx 1.9$~e$^-$/DN.  We dithered by
$\approx 20''$ between each exposure to minimize the effects of bad
pixels and to enable the construction of a super-skyflat from the
unregistered images.  This reduced the effective field of view 
by $\approx 1'$ in each dimension providing $30 - 35 \, \square'$ of 
coverage at full exposure for each field.

We reduced the data following standard techniques.
To produce a super-skyflat for each filter for each night from 
the unregistered images we: (1)
Overscan subtracted each image with an IRAF task specifically designed
to account for the dual-amp readout; 
(2) Masked the frames to remove bad pixels and extended bright objects;  
(3) Scaled each image by the median of the left-hand side and then
implemented the IRAF task {\sc combine} to median the
images using the rejection algorithm {\it ccdclip} with $3 \sigma$ thresholds;
and
(4) Normalized the super-skyflat for each filter
by dividing by the median sky level of the left-hand side of each frame.
The resulting super-skyflats account for both variations in the
pixel to pixel response 
and the illumination pattern of the sky in each filter.
In general, this approach is more advantageous than twilight or dome flats
because it genuinely reproduces the color and illumination pattern of the sky.
The final flattened images are uniform to better than $1\%$ except for
the PSS1443+27 field which exhibits a modest gradient from corner
to corner in the $B$ and $R$ images.

The final stacked images were produced by 
subtracting the overscan and flattening each object image with the
appropriate super-sky flat.  We then measured integer 
offsets\footnote{The seeing did not warrant a more advanced approach like
block replication.} by cross-correlating $\approx 500$ objects identified
by the software package SExtractor.  Finally, we combined the registered images
of each field with an in-house package similar to the IRAF task {\sc COMBINE}
which weights each image by the measured variance, scales each image
by exposure time, and produces a sigma
image based on Poissonian statistics.
The resultant images were then
trimmed by a few rows and columns to register the $BRI$ frames.
We present the images in 
Plates~\ref{fig:0132}-\ref{fig:1443}.
In the following analysis we focus on the regions which received full
exposure, but we also acquired spectra of a few candidates which fell
outside. In the $R$~images,
the objects identified as $B$~dropout candidates are marked with black
circles, the quasar is marked with a large black square, and the 
smaller squares mark the objects with follow-up spectroscopy (Paper~II).
These reduced images are publicly available at
http://kingpin.ucsd.edu/$\sim$dlaimg.  

\begin{figure}[ht]
\caption{$BRI$ images of the field \qoo.  In the images we mark
the $B$-band dropouts with black circles, the quasar with a big black 
box, and the objects selected for follow-up spectroscopy with
black squares.   The images are roughly $5' \times 7'$ and the size
of each pixel is 0.213$''$.  
The images are oriented with S up and E right.
[Postscript versions of these
figures can be downloaded at http://kingpin.ucsd.edu/$\sim$dlaimg]
}
\label{fig:0132}
\end{figure}

\begin{figure}[ht]
\caption{$BRI$ images of the field \qon.  Objects 
are marked with the same scheme as Figure 1.
The images are oriented with S up and E right.
[Postscript versions of these
figures can be downloaded at http://kingpin.ucsd.edu/$\sim$dlaimg]
}
\label{fig:0951}
\end{figure}

\begin{figure}[ht]
\caption{$BRI$ images of the field \qof.  Objects
are marked with the same scheme as Figure 1.
The images are oriented with E up and N right.
[Postscript versions of these
figures can be downloaded at http://kingpin.ucsd.edu/$\sim$dlaimg]
}
\label{fig:1443}
\end{figure}

\section{PHOTOMETRIC CALIBRATION}
\label{sec-photcalib}

During each of the two nights we observed a set of \cite{landolt92}
standard stars in $BRI$ to establish a photometric calibration. We processed
the photometric frames in the same manner as the object frames and 
performed aperture photometry with the IRAF task PHOT.  For the
photometric solution, we adopted 7$''$ radius apertures which matches the
typical aperture employed by \cite{landolt92}. Finally, we performed a 
three parameter fit to each night's photometric data with the IRAF task
FITPARAMS to solve:  

\begin{equation}
M = m - x_1 - x_2 \cdot X_{AM} - x_3 \cdot clr
\end{equation}
where $x_1$ is the zero-point magnitude on a scale where 
25mag yields 1~DN/s, 
$x_2$ is the airmass term and $x_3$ is the color coefficient.  
The color term ($clr$)
refers to $(B-R)_{JC}$ for the $B$ and $R$ solutions and
$(R-I)_{JC}$ for the $I$ measurements.
Table~\ref{tab:photcalib} presents the best fit and errors on
each parameter and all of these values are on the Johnson-Cousins magnitude
scale as appropriate for the Landolt database.  
Because the standard star observations for night~1 did 
not span a significant range of airmass, there is a degeneracy
between the airmass and zero-point terms.  This implies a large uncertainty
in each parameter even though the resulting fit is an excellent match to
the standard star observations.  We chose to minimize this uncertainty by
imposing an airmass term for each filter with values taken from a previous
photometric night of observing with LRIS.  Because the observations of 
{\qon} and {\qoo} were taken at an airmass similar to the standard
stars, small errors in the adopted airmass term will imply small errors
in the derived magnitudes.  
A similar approach was adopted for night~2
where owing to a competing program only two standard star fields were
observed.  In this case, the two standard fields were observed at
significantly different airmass.  This enabled a reasonable fit in the
airmass terms yet the covariance in $x_1$ and $x_2$ was still large enough
to motivate us to adopt the best fit airmass term 
and fit separately for the zero-point.

\begin{table}[ht]
\begin{center}
\caption{ {\sc PHOTOMETRIC CALIBRATION \label{tab:photcalib}}}
\begin{tabular}{ccccc}
\tableline
\tableline
Night & Filter & $x_1$ & $x_2$\tablenotemark{a} & $x_3$ \\
\tableline
Jan 13 1999 & B & $-2.58  \pm 0.02$ & 0.20 & $-0.02 \pm 0.02$ \\
            & R & $-2.59  \pm 0.01$ & 0.08 & $\,\;\;0.05 \pm 0.01$ \\
            & I & $-2.40  \pm 0.02$ & 0.03 & $-0.09 \pm 0.06$ \\
Mar 23 1999 & B & $-2.50  \pm 0.04$ & 0.17 & $-0.02  \pm 0.03$ \\
            & R & $-2.45  \pm 0.01$ & 0.05 & $\,\;\;0.01 \pm 0.01$ \\
            & I & $-2.30  \pm 0.03$ & 0.02 & $-0.19 \pm 0.09$ \\
\tableline
\end{tabular}
\end{center}
\tablenotetext{a}{To lift the covariance between $x_1$ and $x_2$, we
chose to adopt a value for $x_2$ based on our experience or the
best fit value from that night.}
\tablecomments{All of these values are on the Johnson-Cousins magnitude \\
scale \citep{bessel83}.}
\end{table}

The final values assumed throughout the following analysis
are listed in Table~\ref{tab:photcalib}.  Note the relatively large
$(R-I)_{JC}$ color term derived for both nights.  
We expect this term arises because of the non-standard transmission function
of the LRIS $I$ filter.  For most of the objects $|R-I| < 2$,
however, and the color correction is not large.
Comparing the two photometric solutions, we find the zero-point values
for night~2 are systematically $\approx 0.1$~mag higher than for night~1.
We expect the difference is the result of mild extinction (e.g. thin
cirrus clouds) on night~2.  In any case, the offset does 
not have a large effect on our analysis.
Finally, we accounted for Galactic extinction 
by adopting $E(B-V)$ values for each science field taken from 
\cite{schl98} as measured from far-IR dust emission maps:
$E(B-V) = 0.06$ for {\qoo}, 
$E(B-V) = 0.04$ for {\qon}, and
$E(B-V) = 0.03$ for {\qof}.
We assumed a standard Galactic extinction curve with 
$R_V \equiv A(V) / E(B-V) = 3.1$ and applied a correction to the zero-point
magnitudes accordingly.  

\begin{table*}\footnotesize
\begin{center}
\caption{ {\sc PHOTOMETRY FOR PSS0132+13\label{tab:PSS0132+13}}}
\begin{tabular}{lccccrcrcrccccc}
\tableline
\tableline
Id & $x_{pix}$& $y_{pix}$ & RA & Dec & $B_{AB}$\tablenotemark{a} & $\sigma(B)$ & 
$R_{AB}$ & $\sigma(R)$ & $I_{AB}$ & $\sigma(I)$ & 
$z_{phot}$ & Tmpl\tablenotemark{b}& $P_{z}$\tablenotemark{c}& $P_{z \sim 4}$\tablenotemark{d} \\
\tableline
 501& 303.0&1819.8& 23.014034&$+13.657072$&$26.73$& 0.15&25.41& 0.10&$24.37$& 0.16&0.81& 2&0.76&0.00 \\  
 502& 303.1&2126.1& 23.014043&$+13.638950$&$<29.12$& 9.99&26.91& 0.23&$<28.09$& 9.99&3.74& 7&0.41&0.59 \\  
 503& 304.8& 479.5& 23.014138&$+13.736372$&$27.94$& 0.23&26.96& 0.18&$26.25$& 0.17&1.87& 2&0.24&0.02 \\  
 504& 304.9& 687.7& 23.014142&$+13.724052$&$27.70$& 0.22&26.46& 0.13&$25.91$& 0.15&2.42& 2&0.21&0.05 \\  
 505& 305.1& 378.7& 23.014153&$+13.742335$&$26.82$& 0.14&25.90& 0.11&$25.54$& 0.14&0.17& 3&0.26&0.01 \\  
 506& 305.5&1191.6& 23.014185&$+13.694240$&$25.66$& 0.08&24.41& 0.05&$24.00$& 0.07&0.35& 3&0.60&0.02 \\  
 507& 306.0& 311.4& 23.014208&$+13.746318$&$26.89$& 0.21&24.97& 0.07&$24.22$& 0.09&0.59& 2&0.50&0.02 \\  
 508& 306.3&2083.1& 23.014236&$+13.641494$&$<27.15$& 9.99&25.09& 0.09&$23.82$& 0.11&4.70& 5&0.60&0.02 \\  
 509& 306.3&1273.0& 23.014235&$+13.689422$&$25.48$& 0.07&24.30& 0.04&$24.09$& 0.07&3.45& 6&0.76&0.09 \\  
 510& 306.5&1466.5& 23.014244&$+13.677975$&$25.82$& 0.09&24.75& 0.06&$24.77$& 0.11&3.39& 5&0.59&0.02 \\  
 511& 307.1&  41.7& 23.014273&$+13.762278$&$24.56$& 0.07&23.17& 0.03&$22.47$& 0.07&0.06& 1&0.27&0.00 \\  
 512& 308.6&1262.8& 23.014374&$+13.690029$&$26.50$& 0.13&25.22& 0.07&$24.80$& 0.10&0.41& 3&0.43&0.05 \\  
 513& 308.6&1398.9& 23.014375&$+13.681974$&$25.32$& 0.06&24.75& 0.05&$24.36$& 0.08&1.92& 3&0.46&0.00 \\  
 514& 309.1& 395.9& 23.014399&$+13.741320$&$25.41$& 0.07&24.48& 0.05&$24.19$& 0.08&0.16& 3&0.26&0.00 \\  
 515& 311.8& 661.1& 23.014561&$+13.725626$&$26.63$& 0.11&26.05& 0.11&$25.28$& 0.12&1.20& 3&0.51&0.00 \\  
 516& 311.9&1126.5& 23.014572&$+13.698094$&$25.59$& 0.09&24.07& 0.04&$23.09$& 0.08&0.73& 2&1.00&0.00 \\  
 517& 311.9&1051.9& 23.014573&$+13.702504$&$25.81$& 0.07&24.93& 0.06&$24.34$& 0.08&0.09& 2&0.29&0.00 \\  
 518& 312.4& 589.3& 23.014598&$+13.729877$&$27.65$& 0.22&26.47& 0.14&$25.70$& 0.13&0.67& 3&0.33&0.01 \\  
 519& 313.1&1717.4& 23.014650&$+13.663129$&$23.71$& 0.04&22.41& 0.02&$21.73$& 0.07&0.58& 3&0.76&0.00 \\  
 520& 314.0&1035.8& 23.014702&$+13.703462$&$27.35$& 0.21&25.91& 0.11&$25.75$& 0.17&3.45& 7&0.49&0.36 \\  
 521& 314.4&1565.4& 23.014730&$+13.672125$&$<28.17$& 9.99&26.68& 0.15&$<26.91$& 9.99&3.75& 7&0.29&0.52 \\  
 522& 315.7&1043.6& 23.014805&$+13.702996$&$27.33$& 0.26&25.18& 0.08&$24.28$& 0.09&0.41& 1&0.93&0.02 \\  
 523& 315.9&1315.6& 23.014816&$+13.686904$&$25.17$& 0.09&22.97& 0.03&$21.91$& 0.09&0.45& 1&1.00&0.00 \\  
 524& 316.1&1547.0& 23.014829&$+13.673211$&$24.64$& 0.05&23.60& 0.03&$23.26$& 0.06&0.21& 3&0.88&0.00 \\  
 525& 317.7&1302.6& 23.014927&$+13.687676$&$27.30$& 0.19&26.08& 0.11&$25.42$& 0.12&0.61& 3&0.30&0.01 \\  
 526& 317.8& 551.7& 23.014928&$+13.732103$&$28.05$& 0.34&26.18& 0.12&$25.44$& 0.12&0.59& 2&0.48&0.11 \\  
 527& 318.3&1402.3& 23.014965&$+13.681771$&$26.15$& 0.09&25.45& 0.08&$25.20$& 0.12&2.13& 3&0.25&0.00 \\  
 528& 318.6& 796.0& 23.014979&$+13.717645$&$26.16$& 0.08&25.73& 0.09&$25.45$& 0.13&0.65& 5&0.19&0.00 \\  
 529& 318.7&1420.8& 23.014988&$+13.680683$&$26.68$& 0.19&23.80& 0.03&$22.61$& 0.10&0.58& 1&1.00&0.00 \\  
\tableline
\end{tabular}
\end{center}
\tablenotetext{a}{Objects with fluxes $< 3sigma_{sky}$ have reported upper limits equal to the maximum of the flux+1$\sigma_{sky}$ and $1 \sigma_{sky}$ (see $\S$4).
These objects have the error in the magnitude set to 9.99}.
\tablenotetext{b}{Best fit template to the photometry data.  1=Elliptical, 2=Sab, 3=Scd, 4=Irr,
5=SB1, 6=SB2, 7=QSO, $>$7=Stars}
\tablenotetext{c}{Fraction of the integrated likelihood function within 0.2 of $z_{phot}$}
\tablenotetext{d}{Fraction of the integrated likelihood function within the interval $z$ = 3.5 to 4.5.}
\tablecomments{The complete version of this table is in the electronic edition of the Journal.  The printed edition contains only a sample.}
\end{table*}

\begin{table*}\footnotesize
\begin{center}
\caption{ {\sc PHOTOMETRY FOR BR0951--04\label{tab:BR0951--04}}}
\begin{tabular}{lccccrcrcrccccc}
\tableline
\tableline
Id & $x_{pix}$& $y_{pix}$ & RA & Dec & $B_{AB}$\tablenotemark{a} & $\sigma(B)$ & 
$R_{AB}$ & $\sigma(R)$ & $I_{AB}$ & $\sigma(I)$ & 
$z_{phot}$ & Tmpl\tablenotemark{b}& $P_{z}$\tablenotemark{c}& $P_{z \sim 4}$\tablenotemark{d} \\
\tableline
 501& 230.9& 110.9&148.457732&$ -5.032367$&$23.43$& 0.03&22.83& 0.02&$22.85$& 0.04&2.93& 5&1.00&0.00 \\  
 502& 231.5&1076.9&148.457764&$ -5.089522$&$27.94$& 0.18&26.92& 0.15&$<27.44$& 9.99&3.32& 5&0.29&0.03 \\  
 503& 231.5& 940.1&148.457768&$ -5.081425$&$27.32$& 0.14&25.98& 0.09&$25.90$& 0.16&3.55& 5&0.60&0.33 \\  
 504& 231.7&1672.6&148.457779&$ -5.124767$&$26.80$& 0.07&26.65& 0.12&$<26.78$& 9.99&1.20& 7&0.18&0.00 \\  
 505& 231.9& 128.9&148.457790&$ -5.033430$&$26.43$& 0.07&25.30& 0.06&$24.74$& 0.08&2.22& 2&0.59&0.00 \\  
 506& 232.9&1757.4&148.457847&$ -5.129780$&$24.19$& 0.03&23.82& 0.03&$23.52$& 0.05&0.86& 5&0.37&0.00 \\  
 507& 233.1&1918.1&148.457859&$ -5.139288$&$<27.97$& 9.99&25.43& 0.08&$25.51$& 0.15&3.77& 7&0.87&0.98 \\  
 508& 233.3& 889.5&148.457871&$ -5.078428$&$25.48$& 0.04&25.22& 0.05&$25.00$& 0.09&0.06& 5&0.12&0.00 \\  
 509& 233.6& 970.0&148.457889&$ -5.083196$&$25.39$& 0.04&25.09& 0.05&$24.87$& 0.11&2.30& 6&0.23&0.00 \\  
 510& 233.6& 140.2&148.457896&$ -5.034099$&$27.14$& 0.12&26.31& 0.12&$25.77$& 0.14&0.07& 2&0.25&0.00 \\  
 511& 234.1&1054.8&148.457918&$ -5.088212$&$25.32$& 0.05&24.24& 0.03&$23.72$& 0.06&2.12& 2&0.79&0.00 \\  
 512& 234.1&1784.4&148.457919&$ -5.131378$&$28.61$& 0.35&26.89& 0.16&$<26.86$& 9.99&3.70& 5&0.35&0.42 \\  
 513& 234.1& 305.0&148.457923&$ -5.043847$&$27.21$& 0.15&25.66& 0.08&$25.48$& 0.12&3.51& 7&0.73&0.63 \\  
 514& 234.9& 664.2&148.457968&$ -5.065100$&$<28.81$& 9.99&26.71& 0.13&$27.05$& 0.33&3.74& 7&0.65&0.82 \\  
 515& 235.0& 101.6&148.457976&$ -5.031814$&$25.32$& 0.04&24.77& 0.05&$24.86$& 0.10&2.83& 7&0.54&0.00 \\  
 516& 235.4& 701.8&148.458001&$ -5.067323$&$<28.51$& 9.99&27.13& 0.17&$<27.18$& 9.99&3.76& 7&0.24&0.46 \\  
 517& 236.1&1113.3&148.458038&$ -5.091670$&$25.50$& 0.06&24.31& 0.03&$24.38$& 0.07&3.45& 5&0.87&0.08 \\  
 518& 236.1&1640.2&148.458038&$ -5.122845$&$25.69$& 0.07&24.23& 0.03&$23.18$& 0.09&0.77& 2&1.00&0.00 \\  
 519& 236.9&1808.9&148.458085&$ -5.132827$&$23.82$& 0.04&22.82& 0.02&$22.71$& 0.04&3.15& 7&0.65&0.00 \\  
 520& 236.9&2020.8&148.458084&$ -5.145364$&$25.21$& 0.05&24.69& 0.06&$24.50$& 0.09&0.23& 6&0.21&0.00 \\  
 521& 237.6&1305.1&148.458126&$ -5.103019$&$25.55$& 0.09&22.79& 0.02&$21.80$& 0.09&0.53& 1&0.95&0.05 \\  
 522& 237.6& 931.9&148.458128&$ -5.080939$&$27.93$& 0.19&26.35& 0.10&$26.37$& 0.20&3.64& 5&0.68&0.59 \\  
 523& 239.4& 875.2&148.458238&$ -5.077585$&$26.29$& 0.06&25.79& 0.08&$25.65$& 0.13&2.25& 4&0.21&0.00 \\  
 524& 239.7&1632.4&148.458253&$ -5.122386$&$<28.68$& 9.99&26.43& 0.12&$26.70$& 0.29&3.76& 7&0.70&0.89 \\  
 525& 239.9& 344.9&148.458270&$ -5.046206$&$27.13$& 0.12&25.84& 0.08&$25.81$& 0.14&3.52& 5&0.68&0.31 \\  
 526& 240.0& 583.5&148.458270&$ -5.060329$&$24.95$& 0.04&24.44& 0.04&$24.35$& 0.07&2.85& 5&0.27&0.00 \\  
 527& 240.8&   1.7&148.458320&$ -5.025901$&$27.50$& 0.17&26.49& 0.14&$<26.24$& 9.99&3.36& 5&0.21&0.03 \\  
 528& 240.8&1039.7&148.458321&$ -5.087317$&$<29.97$& 9.99&27.04& 0.19&$26.11$& 0.14&4.56& 7&0.59&0.44 \\  
 529& 241.0&1748.3&148.458330&$ -5.129243$&$25.07$& 0.04&24.47& 0.03&$24.46$& 0.07&2.93& 5&0.72&0.00 \\  
\tableline
\end{tabular}
\end{center}
\tablenotetext{a}{Objects with fluxes $< 3sigma_{sky}$ have reported upper limits equal to the maximum of the flux+1$\sigma_{sky}$ and $1 \sigma_{sky}$ (see $\S$4).
These objects have the error in the magnitude set to 9.99}.
\tablenotetext{b}{Best fit template to the photometry data.  1=Elliptical, 2=Sab, 3=Scd, 4=Irr,
5=SB1, 6=SB2, 7=QSO, $>$7=Stars}
\tablenotetext{c}{Fraction of the integrated likelihood function within 0.2 of $z_{phot}$}
\tablenotetext{d}{Fraction of the integrated likelihood function within the interval $z$ = 3.5 to 4.5.}
\tablecomments{The complete version of this table is in the electronic edition of the Journal.  The printed edition contains only a sample.}
\end{table*}

\begin{table*}\footnotesize
\begin{center}
\caption{ {\sc PHOTOMETRY FOR PSS1443+27\label{tab:PSS1443+27}}}
\begin{tabular}{lccccrcrcrccccc}
\tableline
\tableline
Id & $x_{pix}$& $y_{pix}$ & RA & Dec & $B_{AB}$\tablenotemark{a} & $\sigma(B)$ & 
$R_{AB}$ & $\sigma(R)$ & $I_{AB}$ & $\sigma(I)$ & 
$z_{phot}$ & Tmpl\tablenotemark{b}& $P_{z}$\tablenotemark{c}& $P_{z \sim 4}$\tablenotemark{d} \\
\tableline
 501& 265.3&1863.5&220.859171&$+27.359735$&$23.24$& 0.05&22.60& 0.02&$22.28$& 0.07&2.01& 3&0.44&0.00 \\  
 502& 265.6&1819.1&220.859186&$+27.362363$&$25.23$& 0.05&25.08& 0.06&$24.85$& 0.11&1.86& 5&0.31&0.00 \\  
 503& 266.5&  97.7&220.859229&$+27.464213$&$24.71$& 0.07&23.42& 0.03&$22.95$& 0.08&0.22& 2&0.33&0.00 \\  
 504& 268.0&1513.4&220.859344&$+27.380452$&$25.73$& 0.06&25.41& 0.08&$25.74$& 0.25&1.09& 7&0.32&0.00 \\  
 505& 268.7& 716.8&220.859382&$+27.427585$&$27.38$& 0.17&25.60& 0.07&$25.59$& 0.17&3.72& 5&0.87&0.94 \\  
 506& 268.7&1853.3&220.859395&$+27.360342$&$25.54$& 0.06&24.84& 0.05&$24.33$& 0.10&0.68& 4&0.58&0.00 \\  
 507& 269.0& 691.4&220.859403&$+27.429088$&$23.44$& 0.08&21.59& 0.02&$20.99$& 0.09&0.15& 1&0.32&0.00 \\  
 508& 269.2&1507.2&220.859425&$+27.380819$&$28.34$& 0.32&26.48& 0.13&$<27.32$& 9.99&3.69& 5&0.64&0.70 \\  
 509& 270.3&1733.1&220.859499&$+27.367452$&$24.22$& 0.05&23.64& 0.03&$23.24$& 0.08&1.89& 3&0.53&0.00 \\  
 510& 270.7&1031.3&220.859522&$+27.408979$&$26.16$& 0.10&24.78& 0.06&$23.89$& 0.12&0.76& 2&0.80&0.00 \\  
 511& 273.1& 629.5&220.859679&$+27.432750$&$26.12$& 0.09&24.87& 0.05&$24.66$& 0.11&3.48& 6&0.63&0.19 \\  
 512& 273.5&1843.2&220.859714&$+27.360938$&$26.19$& 0.10&25.35& 0.08&$25.09$& 0.15&0.12& 3&0.18&0.00 \\  
 513& 274.2&1162.2&220.859753&$+27.401235$&$27.74$& 0.13&27.31& 0.18&$<27.50$& 9.99&2.72& 7&0.15&0.00 \\  
 514& 275.0&1808.5&220.859815&$+27.362995$&$25.78$& 0.06&25.46& 0.07&$25.39$& 0.17&2.64& 7&0.20&0.00 \\  
 515& 276.1&1319.7&220.859884&$+27.391915$&$27.55$& 0.16&26.25& 0.11&$26.47$& 0.32&3.50& 5&0.54&0.26 \\  
 516& 277.0&1509.5&220.859947&$+27.380684$&$26.48$& 0.09&26.07& 0.11&$25.63$& 0.17&0.90& 6&0.20&0.00 \\  
 517& 277.6& 341.8&220.859972&$+27.449774$&$27.23$& 0.20&25.19& 0.07&$24.11$& 0.14&0.48& 1&0.87&0.01 \\  
 518& 278.2& 814.8&220.860020&$+27.421789$&$<28.90$& 9.99&27.31& 0.19&$<27.27$& 9.99&3.78& 7&0.25&0.51 \\  
 519& 279.4&1970.2&220.860108&$+27.353426$&$24.00$& 0.07&22.36& 0.02&$22.06$& 0.06&3.67& 6&0.98&0.98 \\  
 520& 279.7&1848.2&220.860128&$+27.360644$&$27.06$& 0.13&25.89& 0.09&$25.89$& 0.22&3.44& 5&0.48&0.12 \\  
 521& 279.9&1079.9&220.860134&$+27.406100$&$<28.92$& 9.99&27.35& 0.20&$<27.46$& 9.99&3.76& 7&0.26&0.51 \\  
 522& 280.6&1352.5&220.860185&$+27.389972$&$23.24$& 0.06&22.16& 0.01&$22.00$& 0.05&3.21& 7&0.84&0.01 \\  
 523& 281.1&1619.5&220.860218&$+27.374174$&$25.88$& 0.06&25.57& 0.07&$25.21$& 0.13&1.23& 5&0.24&0.00 \\  
 524& 282.0& 411.1&220.860268&$+27.445674$&$22.17$& 0.07&20.50& 0.01&$20.11$& 0.07&0.35& 2&0.77&0.21 \\  
 525& 282.5&1384.5&220.860310&$+27.388080$&$26.32$& 0.07&26.22& 0.11&$25.72$& 0.17&1.35& 4&0.28&0.00 \\  
 526& 284.2&1201.8&220.860421&$+27.398888$&$27.21$& 0.12&26.52& 0.13&$<26.79$& 9.99&2.98& 5&0.20&0.00 \\  
 527& 284.9&1502.1&220.860470&$+27.381122$&$26.34$& 0.10&25.21& 0.06&$24.85$& 0.12&0.24& 3&0.41&0.03 \\  
 528& 285.1&1712.2&220.860486&$+27.368693$&$26.52$& 0.09&26.00& 0.10&$26.26$& 0.30&2.77& 7&0.24&0.00 \\  
 529& 285.2&1804.7&220.860494&$+27.363219$&$27.77$& 0.14&27.18& 0.18&$<26.93$& 9.99&2.85& 7&0.14&0.00 \\  
\tableline
\end{tabular}
\end{center}
\tablenotetext{a}{Objects with fluxes $< 3sigma_{sky}$ have reported upper limits equal to the maximum of the flux+1$\sigma_{sky}$ and $1 \sigma_{sky}$ (see $\S$4).
These objects have the error in the magnitude set to 9.99}.
\tablenotetext{b}{Best fit template to the photometry data.  1=Elliptical, 2=Sab, 3=Scd, 4=Irr,
5=SB1, 6=SB2, 7=QSO, $>$7=Stars}
\tablenotetext{c}{Fraction of the integrated likelihood function within 0.2 of $z_{phot}$}
\tablenotetext{d}{Fraction of the integrated likelihood function within the interval $z$ = 3.5 to 4.5.}
\tablecomments{The complete version of this table is in the electronic edition of the Journal.  The printed edition contains only a sample.}
\end{table*}

Columns 6 and 7 of Table~\ref{tab:obs} give the $1 \sigma$ surface brightness
limit ($SB^{1\sigma}$; mag/$\square ''$) and a 3$\sigma$ surface brightness
limit per seeing FHWM$^2$ in each filter on the $AB$ scale
for every field assuming no Galactic extinction.
We estimated these limits by measuring the {\it rms} fluctuations in the
sky brightness at several positions where no significant object is 
discernible. Examining the values in Table~\ref{tab:obs}, one notes 
we achieved a detection limit of $I_{AB} < 25.5$ at $5 \sigma$ for an 
unresolved object and are sensitive to $(B-R)_{AB} > 2$ for an object with
$R_{AB} = 26$~mag.

\begin{figure*}
\begin{center}
\includegraphics[height=6.0in, width=4.0in,angle=-90]{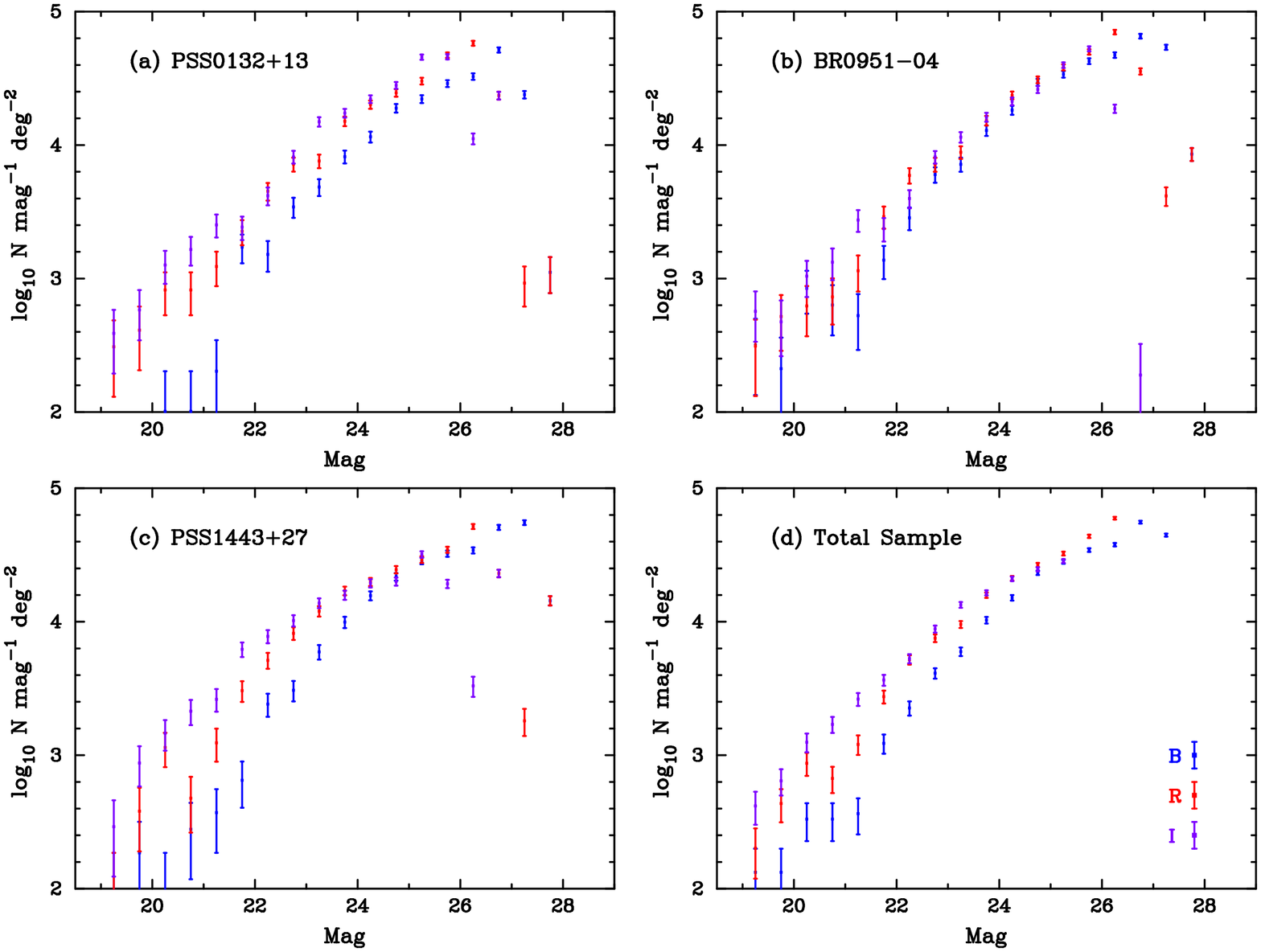}
\caption{$BRI$ number count results for the three fields (a) \qoo, 
(b) \qon, (c) \qof, and (d) the entire sample.  No corrections
have been made for incompleteness although we have displayed only those
counts of the total sample which we expect to have minor completeness
corrections.  Note that all of the magnitudes presented in this 
figure correspond to the Johnson-Cousins system.}
\label{fig:num}
\end{center}
\end{figure*}

\section{PHOTOMETRY AND ASTROMETRY}
\label{sec:photom}

The principal goal of this paper is to pre-select candidates 
at redshifts near $\zabs$ of the damped \lya systems.
We have chosen to survey $z\sim 4$ damped \lya systems, therefore 
these galaxies will be significant $B$-dropouts.  Because
our imaging is more sensitive in $R$ than $I$,
we first identified all candidates in the $R$ images with the software
package SExtractor (v. 2.0.19).  After experimenting with a negative image
of {\qon}, we determined that a 1.1$\sigma$ threshold and 6 pixel minimum
area provided
the largest number of candidates while minimizing spurious detections.
For each field we created a {\sc SEGMENTATION} map with SExtractor
which identifies all of the significant object pixels in the image.
We then used an in-house code to derive the magnitude and error
of each object.
The routine estimates the underlying sky flux by taking the biweight
\citep{beers90}
of the sky in a circular annulus around each object containing at least 1000
pixels not flagged in the {\sc SEGMENTATION} map.   The errors
in the magnitude measurements are derived from variance in the sky and
object flux, and the uncertainty in the photometric calibration.
The latter effect is a systematic error which aside from the color term
is applied in the same way to every object.
Magnitude measurements were repeated for the $B$ and $I$ filters 
restricting the apertures to the same set of pixels identified in the $R$ image.  
We determined that the seeing was sufficiently close between 
the various filters that a more sophisticated analysis was unwarranted.
Many objects of interest are either weakly detected or undetected in the
$B$ images.  If these objects are bright enough for spectroscopy, they
are essentially guaranteed to meet our $(B-R)_{AB} > 2$ criterion.  We
are concerned primarily with the possibility of false positives,
where these weakly detected objects are actually brighter than they appear
due to a negative sky noise fluctuation and do not truly meet the 
$B-R$ criterion.  Therefore,
if the object flux is measured at $< 3 \sigma$ above the sky
flux, we obtain a $1\sigma$ upper limit to the object flux by increasing
its value by $1 \sigma$ sky when 
the object flux is positive or simply setting it to $1 \sigma$ sky 
when the object flux is negative.  Finally, we estimated a corrected
$I$ magnitude for each object using the {\sc MAG\_ISOCOR} routine in
the SExtractor package.  These corrected magnitudes were used to
define our spectroscopic sample of $I_{AB} < 25.5$.
Finally, we performed accurate astrometry on each of our science fields 
to facilitate slit mask design.  We implemented the {\it coordinates}
routine\footnote{ftp://ftp.astro.caltech.edu/pub/palomar/coordinates/coordinates.doc} 
kindly provided by J. Cohen to 
convert our LRIS $(x,y)$ pixel-coordinates to astrometric positions.  We also
identified stars from the USNO and HST Guide star catalogs in our images 
to determine a zero point in RA and DEC.  An optimal USNO/HST star list 
had to be derived taking into account centroid 
inaccuracies of saturated stars and regions of distortion inherent to LRIS.
The instrument's true position angle relative to the sky was obtained using 
the companion routine {\it pa\_check}. 
The position angles for 
PSS0132+1341, BR0951-0450, and PSS1443+2724 were found to be 180.0$\pm 0.46$, 
180.1$\pm 0.25$, and 88.6$\pm 0.56$ degrees respectively.  We estimate the
error in the reported astrometric positions to be $\approx 1.0''$. 

Tables~\ref{tab:PSS0132+13}--\ref{tab:PSS1443+27} summarize the photometry and
astrometry of all of the objects detected in at least one filter for
the three fields.  Columns 1$-$4 list the object positions:
x-pixel, y-pixel, RA, DEC. 
Columns 5--10 the $BRI$ magnitudes and $1 \sigma$ errors,
Finally, column~11 provides the most likely
photometric redshift, column~12 lists the best fit template,
column~13 presents the confidence level
for a $\Delta z = 0.2$ interval centered at the most-likely redshift
($\S$~\ref{subsec:photoz}), and column~14 gives the probability that the
object has $z \; \epsilon \; [3.5,4.5]$.

\begin{figure}[ht]
\includegraphics[height=3.5in, width=3.0in,angle=-90]{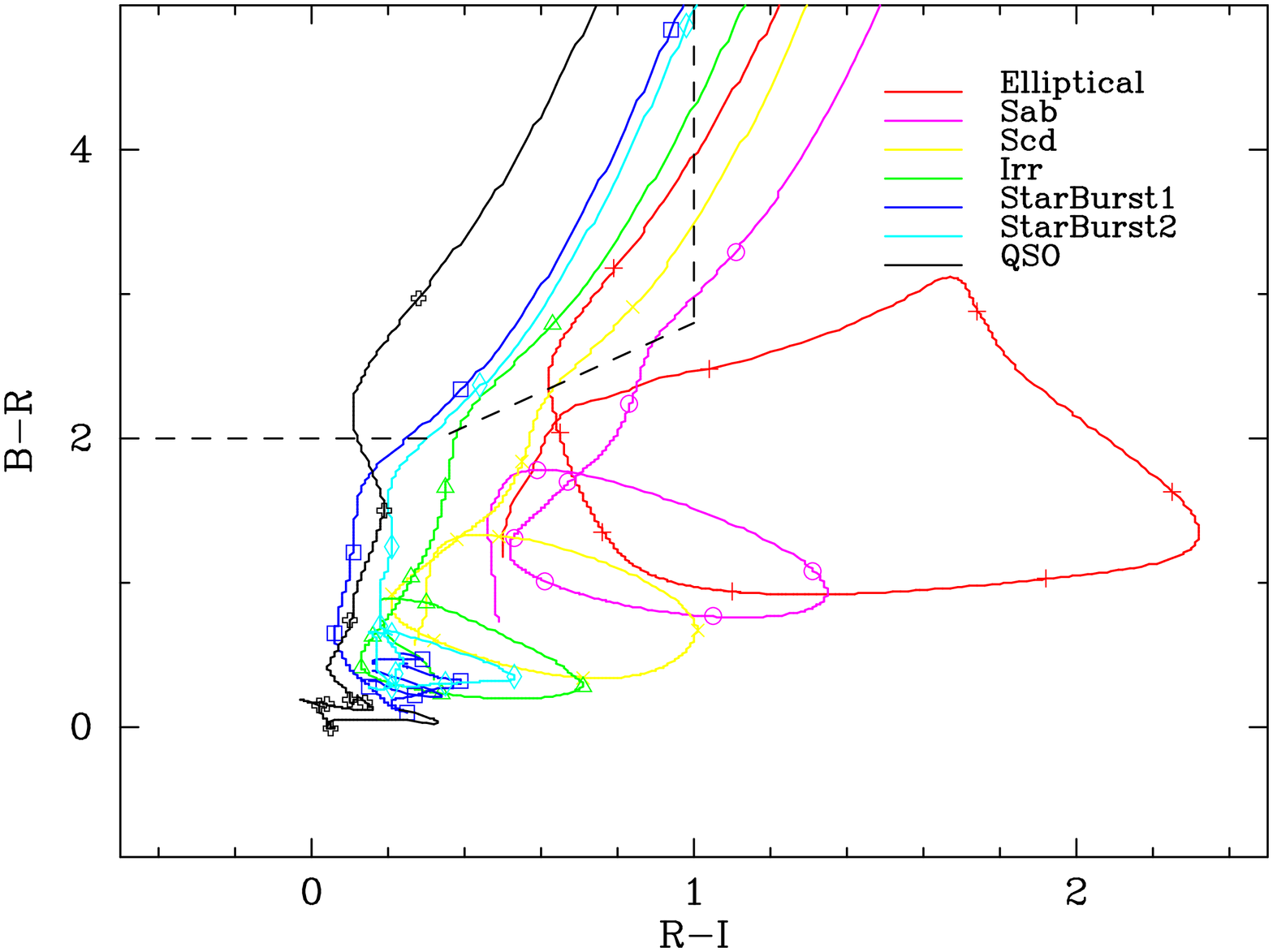}
\caption{Model templates evolved from $z=0-7$ with steps of
$\Delta z = 0.5$ between symbols plotted in a $B-R$ vs.\ $R-I$
color diagram. 
The figure shows that only galaxies with $z>3.5$
lie within the selection region defined by the dashed lines; the
first symbol within the selection region corresponds to $z=4$ for
all templates.  Therefore,
this two color cut is an efficient means of pre-selecting high redshift
galaxies.}
\label{fig:twclrtmp}
\end{figure}

\section{RESULTS}
\label{sec:results}

\subsection{Number Counts}
\label{subsec:num}

We performed a number count analysis\footnote{This
program was not originally designed for a number 
count analysis and these results
should be viewed as tentative.} 
to provide a consistency check on our
photometric calibration and analysis.  We identified objects in each
filter separately with the same approach used to select objects in $R$
for the $B$~drop-out analysis described in $\S$~\ref{sec:photom}, 
i.e.\ the SExtractor package implemented
with a $1.1 \sigma$ threshold and 6 pixel minimum area.
We then measured isophotal and 1.5$''$ diameter aperture magnitudes for each
object and adopted the brighter of the two.
Note that these magnitudes may still underestimate the true magnitude 
as we have not considered isophotal or aperture corrections.
We did not apply a color 
correction to any of the objects,
but we did include the effects of Galactic extinction and airmass.  Finally,
we removed all objects with magnitude brighter than 22  
which were identified by the SExtractor package as a star
(i.e. CLASS\_STAR $> 0.8$).  There may be a small but important
stellar contribution to the counts in the magnitude
$22-22.5$ bin, particularly in the $B$ filter.

Figure~\ref{fig:num} presents the counts for the $BRI$ filters in each
field and for the combined fields.  These are raw counts only; no 
corrections for incompleteness have been applied.  
The magnitudes are on the $AB$ scale, the
error bars assume $\sqrt{N}$ fluctuations and the bin size is 0.5~mag.
While there is some variation from field to field, the differences are
generally small.  
In panel (d) we plot the counts derived by combining the fields.
To crudely account for incompleteness, we conservatively 
truncated the counts in each filter at the magnitude level where there
is a significant 'turn over' (e.g.\ I~$\approx 25.5$ for PSS1443+25).
We find that this criterion corresponds to $\approx 1$~mag brighter than
the $3\sigma SB$ limits presented in Table~\ref{tab:obs}, i.e.\ our
true completeness limit for $z \sim 4$ galaxies is $\approx 1$~mag 
brighter than these $SB^{3\sigma}_{FWHM}$ values.
Examining this panel, we observe results similar to those
described by other number count studies \citep[e.g.][]{steidel93,smail95}.
In particular, the $B$~counts exhibit the steepest rise at $22 - 25$~mag
with evidence for a flattening at fainter magnitudes.  This bend in the
slope is independent of completeness corrections because it occurs
well above our detection limit.  Meanwhile, the $R$ and $I$
slopes are nearly constant to the completeness limit.  We note that the
absolute scale of the number counts is lower than the
\cite{smail95} results and expect the difference is due to the exact
assumptions for apertures and corrected magnitudes. In summary, 
the number count data lend confidence to our photometric calibration
and analysis.

\subsection{Two-color Analysis}
\label{subsec:color}

\cite{steidel99} were the first to demonstrate that one can efficiently
pre-select $z \sim 4$ star-forming galaxies with broad band images.
The approach is to design a set of color criteria which keys on the
sharp flux decrement at the $\lambda \sim (1+z) \times 912$~\rAA Lyman break
while also considering the substantial flux decrement due to absorption
by intergalactic neutral hydrogen.  The approach
is well described by Figure~\ref{fig:twclrtmp} where we plot
$(B-R)_{AB}$ vs.\ $(R-I)_{AB}$ colors for 
six galactic templates and one QSO template (see Yahata et al.\ 2000) 
adopting the quantum
efficiencies of LRIS on the Keck telescope\footnote{Absorption due to 
intergalactic hydrogen was included in our analysis following the
prescription of Madau (1995).}.  
The colored lines trace the evolutionary tracks of each template as
they are $K$-corrected from $z = 0$ to 5 and the points mark each
$\Delta z = 0.5$ interval.  
The dashed line demarcates
the region of color-color space defining our selection criteria designed
to include 
galaxies and the quasar with $z>3.5$ while minimizing the number of
contaminating galaxies:
\begin{eqnarray}
(B-R)_{AB} > 2.0 \\
(R-I)_{AB} < 1.0 \\
(B-R)_{AB} > 1.2 (R-I)_{AB} + 1.6
\end{eqnarray}
These criteria follow the constraints imposed by \cite{steidel99} but were
modified to account for our specific filter set.
The large $B-R$ color criterion
reflects the flux decrement due to the Lyman break
and the \lya absorption from intergalactic HI gas.
In short, 
Figure~\ref{fig:twclrtmp} demonstrates that
by restricting the objects to the upper left-hand corner of this
color-color diagram we are sensitive to all galaxy types in the interval
$3.5 < z < 4.5$ except those with rest-frame UV properties similar to
Sab galaxies (i.e.\ galaxies with minimal ongoing star formation).

\begin{figure*}
\begin{center}
\includegraphics[height=6.0in, width=4.2in,angle=-90]{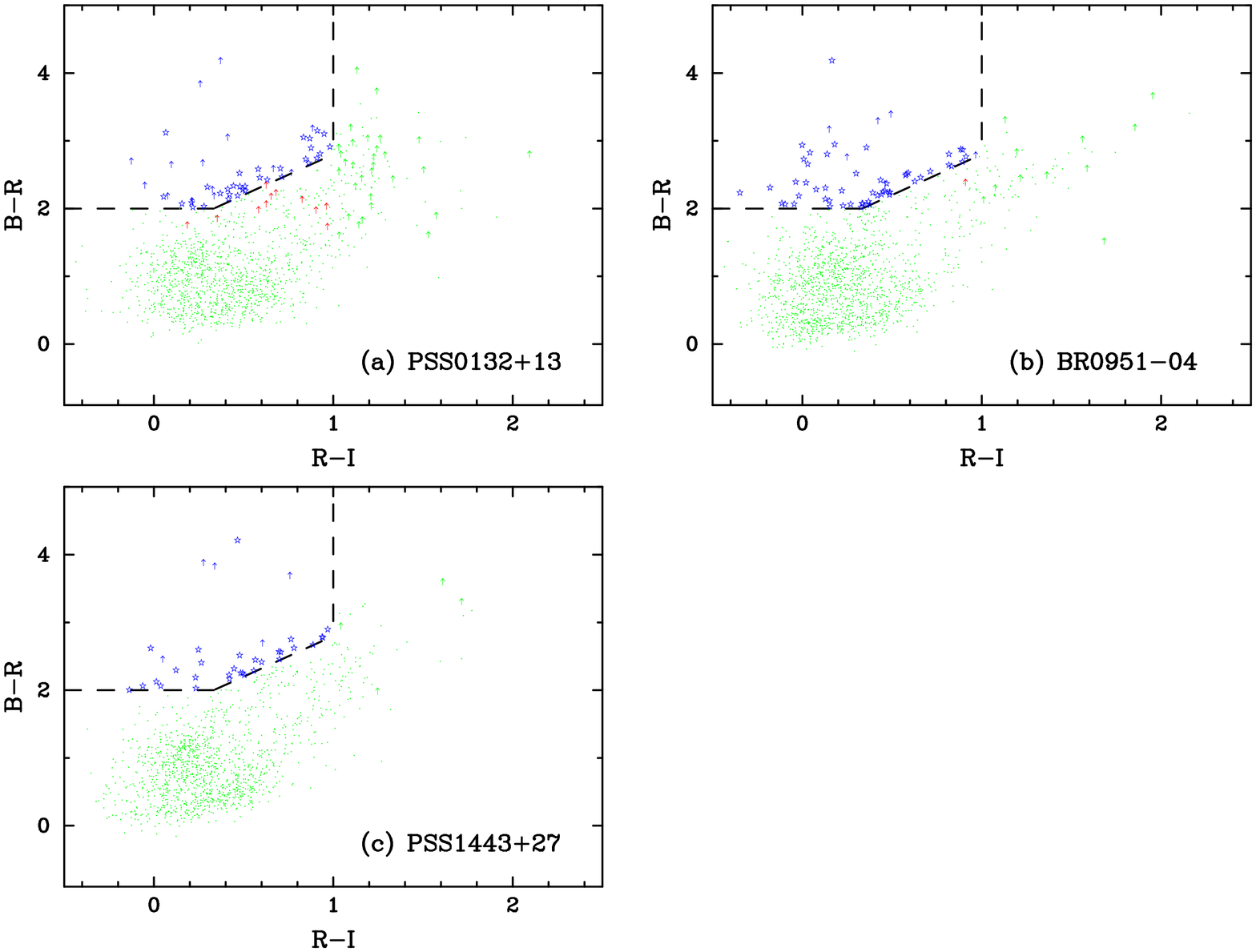}
\caption{Two color diagram of all of the objects with a significant detection
in the $R$~image.  The blue stars and arrows are candidates which satisfy
the selection criteria (designated by the dashed lines), the red
arrows are galaxies with limits to their colors which might 
lie within the selection region, and the green points do not satisfy the
criteria. }
\label{fig:twclr}
\end{center}
\end{figure*}

Figure~\ref{fig:twclr} presents the color-color diagrams for all of the 
objects in the full exposure analysis regions of the three fields.
As discussed in $\S$~\ref{sec:photom},
we selected objects and isophotal apertures 
in the $R$-band and then measured the $BRI$ magnitudes. 
In Figure~\ref{fig:twclr} we have limited the objects to $I_{AB} < 25.5$ 
which corresponds to an (ambitious)
intrinsic spectroscopic limit set by a 2~hr integration
time with LRIS on the Keck telescope.  The dark blue points
indicate objects which satisfy the color criteria, the red objects 
have a measured magnitude limit (typically in $B$) and might
satisfy the criteria, and the green points are excluded from the selection 
region.  In the case where we measure only a limit for a given color,
we plot the point as an arrow.
We visually inspected all candidate galaxies in our
images and rejected roughly half of the initial candidates for having
flawed photometry.  The rejected candidates were typically near bright 
extended objects, satellite trails, or the borders of the images.  It is 
no surprise that such a high fraction of the initial candidates had 
flawed photometry, as the color selection region is sparsely populated 
and therefore the small fraction of low-redshift galaxies scattered into 
the color selection region by photometric errors can easily comprise a
large fraction of the candidates.  Only a small fraction of the true 
high-redshift galaxies should be rejected during this step, although some 
of them may scatter out of the selection region due to photometric 
errors or unusual intrinsic colors.  
Bright (I$<23$) objects with clearly non-stellar morphology were rejected as
low-redshift interlopers, but bright
stellar objects were retained in case they
were new quasars or lensing-magnified LBGs;
most LBGs will look similar to point sources
at the resolution of our images.
The squares identify those candidates which were 
selected for follow-up multi-slit spectroscopy (see Paper~II).
For each field, $\sim 30 - 40$ objects lie within the selection region and
an additional $\sim 10$ limits are consistent with the region.
One can reliably construct a mask with 15$-$20 slits covering roughly half
of the LRIS imaging field.  Because some candidates conflict with others 
(particularly in the {\qon} field),
we had the freedom to choose $\approx 5$ objects per mask 
which lie outside the selection region.  
The majority of these were 
chosen according to their photometric redshift.
This explains why some objects in Figures~\ref{fig:0132}-\ref{fig:1443}
are marked as spectroscopic targets but not color dropouts.

\begin{figure*}
\begin{center}
\includegraphics[height=6.9in,angle=-90]{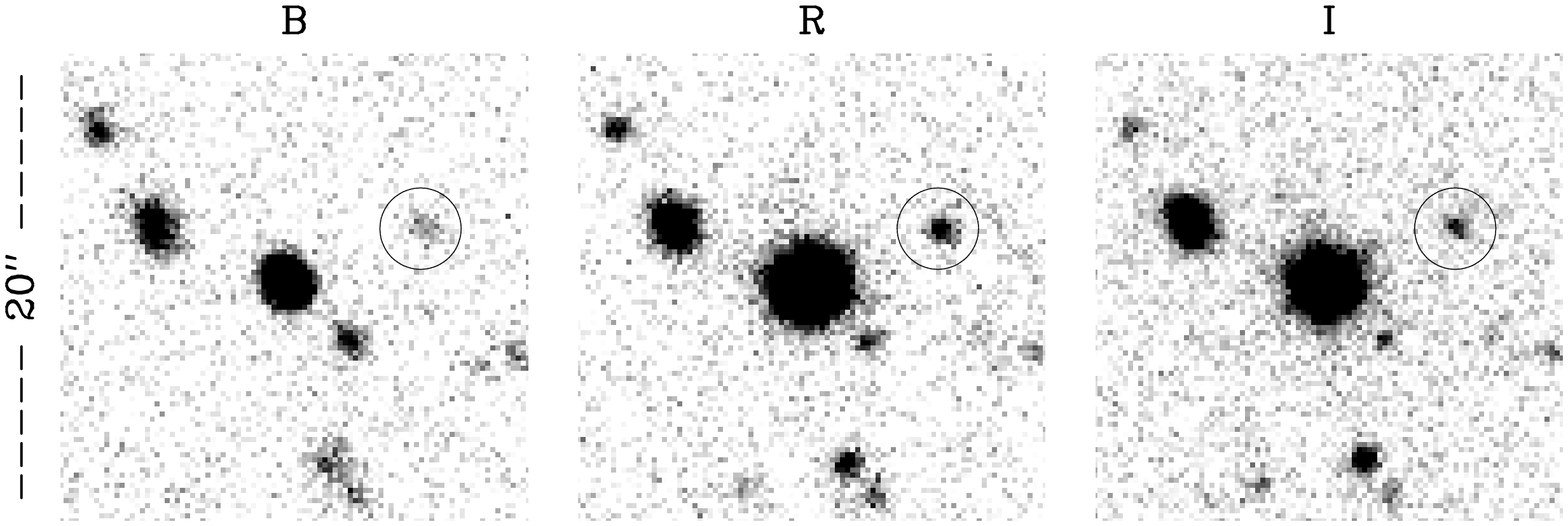}
\caption{Close-up $BRI$ images ($20'' \times 20''$) centered on the
quasar \qoo.  The object identified to the right of the quasar is the
only significant $B$-band dropout in this region and has a photometric
redshift $z_{phot} = 3.6$.  The images are oriented with S up and E right.}
\label{fig:qsoa}
\end{center}
\end{figure*}

\subsection{Photometric Redshifts}
\label{subsec:photoz}

Implementing the procedures introduced by \cite{lanzetta96} and
\cite{nori2000},
we have calculated photometric redshift likelihood functions
for each object.  
The photometric redshift method accounts for the individual
photometric uncertainties in the $B,R,$ and $I$
fluxes of each object and compares the
object fluxes with a set of template spectra rather than a
single two-color region.
We adopted 6 galactic templates (elliptical, Sab, Scd,
irregular, a reddened starburst, and an unreddened starburst), 
1 quasar template,
and $\sim$100 stellar templates with spectral type varying from O to M.
We then stepped the non-stellar templates
in 0.01 redshift intervals from $z = 0$ to 7, 
identified the best fit template for each redshift,
and recorded its $\chi^2$.  Finally, we calculated the 
likelihood of the best fit template at each redshift interval to determine the
most likely $\zphot$ value.  As we are limited to only 3 filters per field,
the confidence with which we can attribute a single redshift to a given 
object is often poor.  For example, an object with $BRI$ magnitudes
consistent with a $z=0.5$ elliptical galaxy is also well matched by a
$z \approx 3.5$ Sab template.  This degeneracy is revealed in 
Figure~\ref{fig:twclrtmp} at $B-R=2$ and $R-I=0.8$.  In other words,
essentially only those galaxies which satisfy the color-color criteria
could be confidently identified as $z \sim 4$ galaxies.  
Therefore, in concordance with previous studies,
we consider the selection region to be the most robust approach
to efficiently identify a significant number of $z \sim 4$ galaxies in our
fields. 

In each of the fields, we identified a number of objects with $z_{phot} \sim 4$
that lie outside our color-color criteria.  The majority of these galaxies
were fit with early type spectra, not starburst models.  We have included
several of these candidates in the spectroscopic program in part to fill
the slit masks (as noted above) but also to test the color-color criteria.
In turn, we will better define our selection function for galaxies at
$z \sim 4$. 
Preliminary results from \cite{gawiser01a} indicate that
with $BRI$ photometry alone, the photometric redshifts of galaxies
confirmed spectroscopically to be at $3.5<z<4.5$ are quite accurate.  However,
a large number of low-redshift objects were aliased to high photometric
redshifts due to color degeneracies between
Lyman-break galaxies and low-redshift ellipticals \citep[e.g.][]{steidel99}. 
The increased fraction of
interlopers among candidates
selected primarily by photometric redshift confirms that the
color selection region modified from
\cite{steidel99} is the most efficient place to find LBGs.

\subsection{The Damped \lya Systems}

\begin{figure*}
\begin{center}
\includegraphics[height=6.9in,angle=-90]{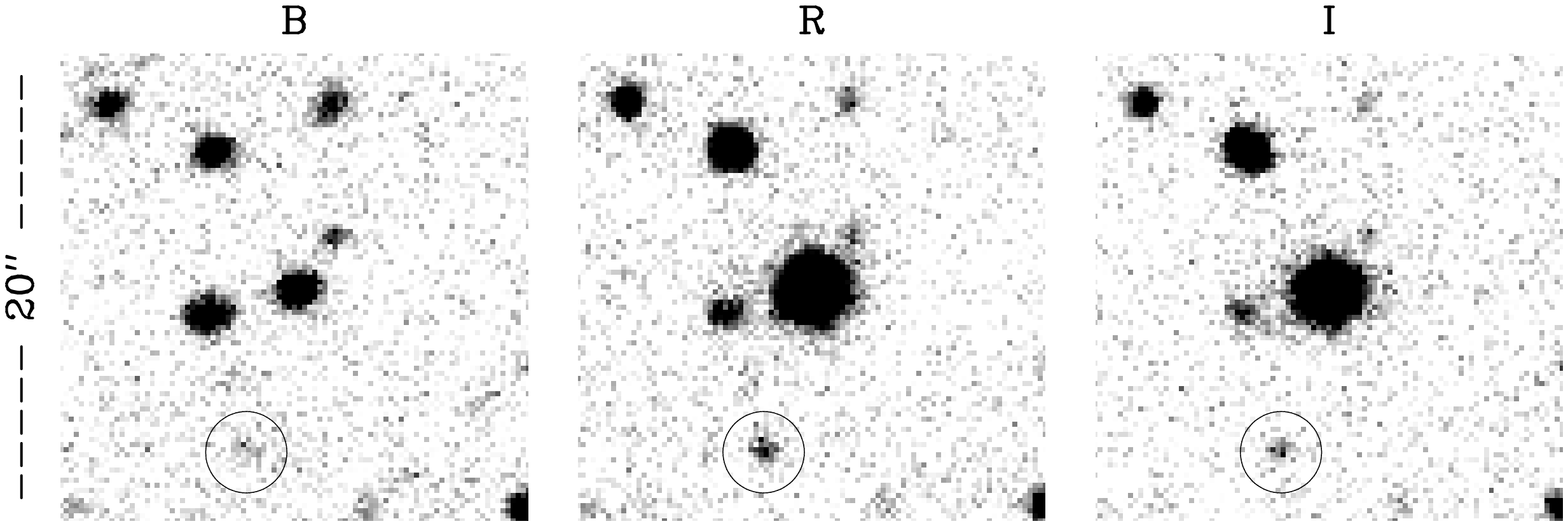}
\caption{Close-up $BRI$ images ($20'' \times 20''$) centered on the
quasar \qon.  The marked object below the quasar is the
only significant $B$-band dropout in this region and has a photometric
redshift $z_{phot} = 3.7$.  The images are oriented with S up and E right.}
\label{fig:qsob}
\end{center}
\end{figure*}

\begin{figure*}
\begin{center}
\includegraphics[height=6.9in,angle=-90]{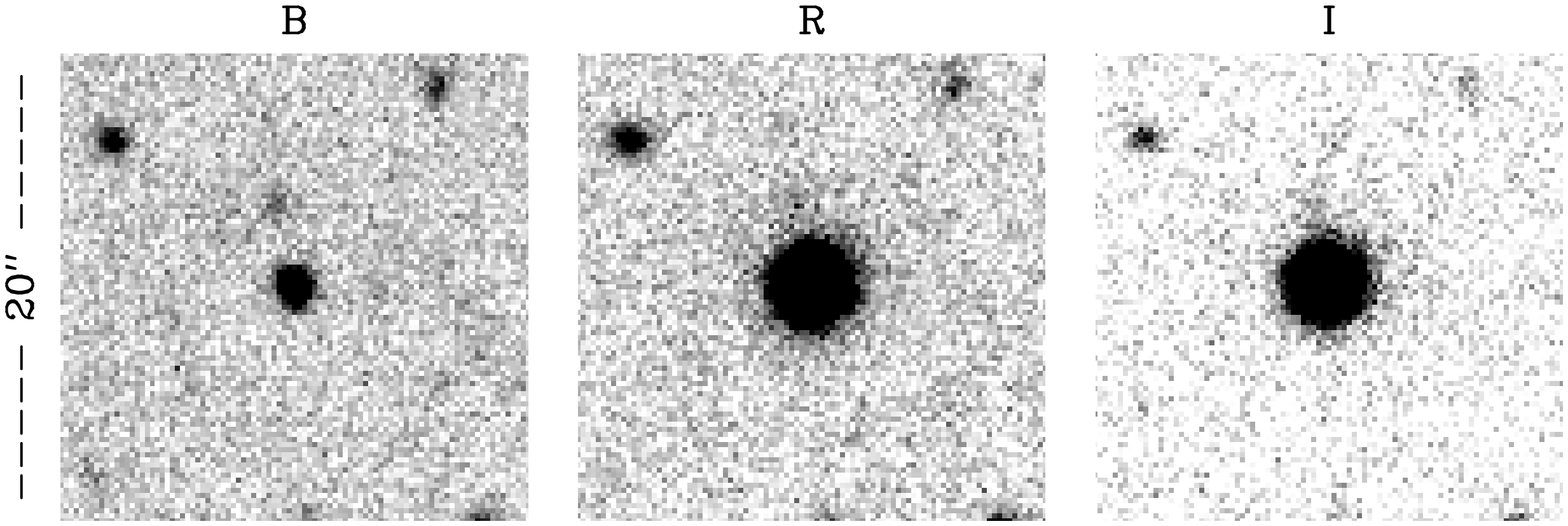}
\caption{Close-up $BRI$ images ($20'' \times 20''$) centered on the
quasar \qof. Unlike the other two fields, there is not a single 
significant $B$-band dropout in this region.  In fact, there are 
an unusually small number of galaxies of any type near the quasar.  
The images are oriented with E up and N right.}
\label{fig:qsoc}
\end{center}
\end{figure*}

Figures~\ref{fig:qsoa}--\ref{fig:qsoc} present 20$'' \times 20''$ 
$BRI$ images centered
on the quasar of each field.  At $z = 4$, a $10''$ radius corresponds to 
65~$h_{75}^{-1}$~kpc ($q_0 = 0.15, H_0 = 75 h_{75}$~km/s/Mpc) 
which we expect
encompasses the center of the galaxy giving rise to the damped
\lya system.  For \qon, there is a marginal $B$~drop-out ($B-R = 1.7$)
$\approx 7''$ below (North) the quasar.  
Its photometric redshift is $z_{phot} = 3.7$
and its impact parameter corresponds to $\approx 50 h^{-1}_{75}$~kpc.  
Although this separation would imply a very large gas disk in order to
explain the observed $\N{HI}$ of either damped system, it can not be ruled out
as the galaxy responsible for the absorption.
There are no other significant $B$~drop-out candidates in this 20$''$ circle.  
Examining the {\qoo} field, we note a marginal $(B-R = 1.5$)
$B$~drop-out object $\approx 7''$ to the right (East)
of {\qoo} with $z_{phot} = 3.6$.
Given the large separation and lower $B-R$ value, it is even 
less likely that this galaxy gives rise to the damped \lya system.
In contrast to the {\qoo} and {\qon} fields, the region surrounding {\qof}
is notably blank.  In fact, there are only three significant galaxies in the
20$'' \times 20''$ area and only an additional 8 
in a $30'' \times 30''$ box, none of which are $B$-band dropouts.  
The absence of candidates is somewhat striking given that this
system has the highest metallicity of any known $z > 3$
damped \lya system and might have been expected to
exhibit an optical counterpart.

We have attempted to subtract the point-spread function (PSF) of
the quasar to search for damped \lya candidates underneath the PSF.
We first used the software package (DAOPHOT) and IRAF routines
to model the PSF of many bright stars throughout each field.
We then focused on several stars close to the QSO to account
for variations in the PSF across the field due to aberrations inherent to
the LRIS imager.  We then fit this PSF model to the quasar and
subtracted it from the image.  Unfortunately, owing to the
segmented nature of the Keck telescope we found it too difficult to
accurately model the PSF and had no success identifying objects 
underneath the PSF.  Studies which take advantage of higher
spatial resolution and a better behaved PSF
(e.g. HST) are far more effective \citep[e.g.][]{warren01}.

In summary, the only possible optical counterparts for the four
damped \lya systems down to $I_{AB} < 25.5$ lie at impact parameter
$\approx 7''$ from the sightline.  This is a significant separation
even without considering projection.  We expect -- if anything -- that these
galaxies are clustered with the damped \lya system.  
If we assume that at least one of the galaxies giving rise to the damped
\lya systems does not lie beneath the PSF of the quasar
\citep{warren01}, then we can place an upper limit
to its luminosity based on the depth of our $I$ images.  At $z=4$,
\cite{steidel99} have fit a Schechter luminosity function to their sample
of Lyman break galaxies and find $L^*_{LBG}$ corresponds to 
$I_{AB} \approx 25$~mag. 
For the \qoo\ and \qof\ fields, where no reasonable candidate was 
identified in the images, our $3 \sigma$ limits in the $I$-band of
27.4 and 26.6 correspond to $L < L_{LBG}^*/9$ and $L_{LBG}^*/4$ respectively.

\section{SUMMARY}
\label{sec:summary}

We have presented an analysis of $BRI$ images for three fields containing 
quasars with four known damped \lya systems at 
$z \sim 4$.  The data was reduced with standard procedures and calibrated
with Landolt standard stars taken close to the time of the observations.
We measured $BRI$ magnitudes for all of the objects in each field
using isophotal apertures and measured $B-R$ and $R-I$ colors for those
galaxies detected in the $R$-images.  A rough number count analysis agrees
well with other studies in the literature to similar magnitude limits.

By searching for $B$-band dropouts and measuring photometric
redshifts, we have culled a list of candidates at $z \sim 4$
throughout each field for follow-up spectroscopy.  We identified 
$\approx 30 - 40$ candidates per field 
to a limiting magnitude of $I_{AB} < 25.5$.  For the three fields there are only
two dropouts detected within $10''$ of the quasar, each at an
impact parameter of $\approx 7''$.
Although these two galaxies could give rise to one or two of the damped
\lya systems, we expect it is more likely that they are either clustered
with the damped system or at a different redshift altogether.  Therefore, 
we have no convincing detections of the galaxy in emission
for four damped systems in the three fields.
Either the galaxies are fainter
than our detection limit ($\approx L_{LBG}^*/4$) or located under the 
PSF of the background quasar.  Nevertheless, our results
enable a study of the large-scale clustering of galaxies associated with
the damped \lya systems.  Indeed,
this is the focus of the analyses presented in companion papers (Paper~II).

\acknowledgments

The authors wish to recognize and acknowledge the very significant cultural
role and reverence that the summit of Mauna Kea has always had within the
indigenous Hawaiian community.  We are most fortunate to have the
opportunity to conduct observations from this mountain.
We acknowledge the very helpful Keck support staff for their assistance
in performing these observations.  We would like to thank R. Bernstein,
B. Weiner, and K. Adelberger for helpful discussions.   
This work was partially supported by NASA through a Hubble Fellowship
grant HF-01142.01-A awarded by STScI to JXP.
AMW and JC were partially supported by NSF grant AST 0071257.

\clearpage

\end{document}